# Designing of Organic Bridging Linkers of Metal-Organic Frameworks for Enhanced Carbon Dioxide Adsorption

Kahkasha Parveen[1] and Srimanta Pakhira[1,2]*

[1] Theoretical Condensed Matter Physics & Advanced Computational Materials Science Laboratory, Department of Physics, Indian Institute of Technology Indore (IIT Indore), Simrol, Khandwa Road, Indore, Madhya Pradesh, 453552, India.

[2] Centre for Advanced Electronics (CAE), Indian Institute of Technology Indore, Simrol, Khandwa Road, Indore, Madhya Pradesh, 453552, India.

**Corresponding author:** spakhira@iiti.ac.in (or) spakhirafsu@gmail.com

**ABSTRACT**

The global rate of anthropogenic carbon dioxide ($CO_2$) emission is rising, which urges the development of efficient carbon capture and storage (CCS) technologies. Among the various $CO_2$ capture methods, adsorption by the linkers of the Metal-Organic Frameworks (MOFs) materials has received more interest as excellent $CO_2$ adsorbents because of their important role in understanding the interaction mechanism for $CO_2$ adsorption. Here, we investigate the adsorption of $CO_2$ molecules at the center and side positions of several MOF-linkers using molecular cluster models. The interaction between $CO_2$ and the linkers is approximated by computing the binding enthalpy ($\Delta H$) through the first principles-based Density Functional Theory with Grimme's dispersion correction (i.e., B3LYP-D3) and second-order Møller Plesset Theory (MP2). The computed values of $\Delta H$ indicate the weak nature of $CO_2$ adsorption on the pristine linkers, hence the strategy of lithium decoration is used to see its impact on the binding strength. Among the various linkers tested, $CO_2$ adsorbing at the side position of the DFBDC-2 linker has strong adsorption with $\Delta H$ value of about -35.32 kJ/mol computed by the B3LYP-D3 method. The Energy Decomposition Analysis (EDA) study reveals that among all the energy terms, the contribution of electrostatic and polarization energy terms to the $\Delta H$ value are the most dominant one. Furthermore, the results of Frontier Molecular Orbital Analysis (FMO) revealed that all the linkers remained stable even after Li-decoration. The results of our investigations will direct towards the development and synthesis of novel adsorbents with enhanced $CO_2$ adsorption.

## 1. INTRODUCTION

The burning of fossil fuels for the purpose of power generation and transportation emits many greenhouse gases in the environment causing pollution, climate change, and harmful impact on human health. Carbon dioxide ($CO_2$) is believed to be one of the main contributors to the greenhouse gases which have triggered serious global climate change.[1] Thus, to reduce the emission of $CO_2$ and minimize the impact of greenhouse gases on the environment, the development of feasible technologies for efficient $CO_2$ capture is highly required.[2] The captured $CO_2$ can be reused as a source of carbon and can be effectively converted into sustainable and value-added fuels such as methane, methanol, formaldehyde, etc.[3,4] Among the various techniques such as chemical absorption, adsorption, cryogenic and membrane separation, carbon capture and storage (CCS) by solid porous adsorbents have evolved as an innovative strategy that can be used to reduce the emission of $CO_2$.[5–7] Adsorption by porous materials presents the advantages of quick recovery, high adsorption capacity under humid environment, simple handling, and material's stability.[8] However, the development of efficient adsorbents with high adsorption capacity, thermal and chemical stability, fast kinetics, and low cost with reversible $CO_2$ uptake and release is the most challenging intent at present. A wide variety of adsorbents such as covalent organic frameworks (COFs),[9] metal-organic frameworks (MOFs),[10,11] and zeolites[12–15] have been employed for storage, capture, and separation of $CO_2$, $H_2$, $CH_4$, $N_2$ from the mixture of gases.

MOFs are a class of crystalline nanoporous materials that have been used for several applications such as gas storage and separation processes catalysis, and sensing.[16–18] The ability of the MOF materials to be synthesized with different bridging linkers/ligands and metal ions/clusters provides enormous flexibility in the design of these porous materials with tunable pore size, pore volume, and high surface area.[19,20] These properties have made MOF materials an interesting topic for the adsorption of gases in modern science and technology, however only a few studies have been done for $CO_2$ capture using molecular dynamics and simulations. Babaroa and Jiang carried out a molecular simulation for $CO_2$ capture in a number of MOFs at ambient temperature, and they found that organic linker are crucial for adjusting the free volume and accessible surface area as well as determining $CO_2$ uptake at high pressure.[21] Torrisi et al. analysed the intermolecular interactions between $CO_2$ molecules and a number of functionalized aromatic molecules using the density functional theory (DFT) method to provide guidance for the design of linkers that can form new materials with enhanced energy for $CO_2$ adsorption.[22,23] Moreover, a study by Liu et al. explained the adsorption mechanism of $CO_2$ on linkers of IRMOF-1 by means of density functional theory. The authors studied eight different positions with three orientations of $CO_2$, suggesting the side position with $CO_2$ parallel attack at hydrogen side of linker edge is the most favourable adsorption site.[24] These studies suggests that interaction energy, positions and orientations of the gas molecule, along with the organic linkers of the MOFs materials instigate to the understanding of $CO_2$ adsorption mechanism and designing of novel MOF with their applications in high uptake quantity. However, due to the low reactivity (interaction energy) of the adsorbate or linkers surface, the adsorption of $CO_2$ on the pristine MOF linkers belongs to the weak physical adsorption, limiting the practical application of MOF as an adsorbent material. According to findings from earlier research, altering the adsorbate surface through techniques like metal doping or decoration, chelation, functionalization, and ligand/linker modification boosts the chemical reactivity of the substrate and strengthens the interaction between the gas molecules and the linker-based materials.[25–32] Thus, the increased strength of interaction with the surface will help in understanding the interaction mechanism of $CO_2$ adsorption and the designing of an ideal adsorbate material for their gas storage applications.

In this work, we have computationally designed eight linkers i.e., 1,4-benzene dicarboxylate (BDC), 3-fluoro 1,4-benzene dicarboxylate (FBDC), 3,6-difluoro 1,4-benzene dicarboxylate (DFBDC-1), 2,3-difluoro 1,4-benzene dicarboxylate (DFBDC-2), 2,3,5,6-tetrafluoro 1,4-benzene dicarboxylate (TFBDC), 3,6-dichloro 1,4-benzene dicarboxylate (DClBDC-1), 2,3-dichloro 1,4-benzene dicarboxylate (DClBDC-2), and 1,4 naphthalene dicarboxylate (NDC) using molecular cluster model systems. Here, we employed first principles-based hybrid periodic density functional theory with van der Waals (vdw) corrections (i.e., Grimmes-D3 dispersion corrections) to calculate the equilibrium geometries of the complexes. The single point energy calculations have been evaluated using the DFT (B3LYP-D3) and second-order Moller Plesset Theory (MP2) methods. The study of binding or relative enthalpy (ΔH) shows the weak interaction of the $CO_2$ molecule with the linkers, so the strategy of Li-decoration has been used to see its effect on the ΔH value. We computationally found that among all the complexes, $CO_2$ adsorbing at the side position of the DFBDC-2 linker have the strongest interactions having ΔH value of -35.32 kJ/mol. Further to explain the interaction mechanism and find the contribution from different energy terms to the ΔH value, the Energy Decomposition Analysis (EDA) was performed, and electrostatic and polarization energy terms were found to be the most dominant one. Also, to check the stability of the complexes, Frontier Molecular Orbital (FMO) analysis was performed for the complexes showing strong interaction and was found that the complexes remained stable even after Li-decoration.

## 2. METHODS AND COMPUTATIONAL DETAILS

The mechanism of $CO_2$ physisorption on the linkers has been investigated by employing the first principles based unrestricted hybrid density functional theory (UDFT) B3LYP method. The equilibrium structures, geometries, energies of the molecular orbital, and the contribution of different energy component to the ΔH value of the complexes were obtained by employing the B3LYP DFT method implemented in Gaussian16 suite code.[33] The strength of interaction of the $CO_2$ molecule with the linkers was evaluated by calculation the single point energy of the complexes performed by using B3LYP DFT method and second-order Moller Plesset Theory (MP2) method. To incorporate the long-range dispersion effect between the atoms of the complexes, Grimme's dispersion corrections parameter (-D3) has been added with the DFT approach (i.e., B3LYP-D3) which is necessary for the weakly bound systems.[34–40] The hybrid B3LYP functional has been widely used for geometry optimization for systems containing metal atoms in their constitution, providing an accurate description and fair indication of the interaction energy.[41] The MP2 method has been widely used for predicting the weak van der Waals (vdW) forces accurately but suffers from the disadvantage of time consumption and its applicability on large systems. So, to reduce computational costs, DFT methods with Grimmes dispersion corrections have been used in most places instead of the high computational cost MP2 method. In comparison to MP2, the DFT-D method offers a good balance between accuracy and computational efforts. For describing the structures and properties of organic linkers as well as metal, it is important to choose an appropriate basis set. Therefore, to perform the calculations like getting the equilibrium geometries and defining the atomic orbitals of all the atoms in the linkers, the correlation consistent polarization valence triple-ζ quality Gaussian basis set (cc-pVTZ) has been utilised.[42] This basis set is advantageous over other basis sets as it contains higher basis function for each atom and gives much better results with high accuracy in the calculations. Using the same methods and basis sets, frequency (i.e., harmonic vibrational analysis) and thermodynamic calculation were performed to obtain the value of ΔH and confirm the stable geometry of the complexes. All the structures were optimized without any symmetry constraints and the optimized minimum-energy structures were confirmed to be stationary points on the potential energy surface. The chemical structure and analysis of the optimal geometry studied were created and visualized using the visualization tool Chemcraft.[43] For the pictorial representation of the adsorption sites, we used GaussView 6.1.1, a graphical interface used with Gaussian.[44]

The calculation of relative or binding enthalpy (ΔH) includes geometry optimizations of both the pristine or pure and Li-decorated linkers with one $CO_2$ getting adsorbed to the center and side positions. After obtaining the equilibrium structure, the values of ΔH were calculated using the same B3LYP-D3 and MP2 methods. To check the performance of both the methods, equations 1 and 2 were used for calculating the values of ΔH for both the pure and Li-decorated case respectively.

$$\Delta H_{CO_2} = H_{(linker + CO_2)} - (H_{linker} + H_{CO_2}) \quad \ldots (1)$$

$$\Delta H_{CO_2} = H_{(linker + Li + CO_2)} - (H_{linker + Li} + H_{CO_2}) \quad \ldots (2)$$

Where $H_{(linker + CO_2)}$ represents the enthalpy of the adsorbate/substrate system in an equilibrium state with $CO_2$ molecule, $H_{linker}$ represents the enthalpy of the linker, $H_{(linker + Li + CO_2)}$ is the enthalpy of the linker after Li decoration/adding with $CO_2$, $H_{linker + Li}$ is the enthalpy of the Li decorated linker and $H_{CO_2}$ is the enthalpy of the $CO_2$ molecule. The negative value of the binding enthalpy indicates that the adsorption of $CO_2$ is exothermic in nature and a higher negative value of ΔH corresponds to a stronger adsorption. Binding enthalpy is an essential factor affecting the MOFs ability to bind $CO_2$ and other gas molecules. Adsorption of gases in the MOF adsorbent would greatly benefit from an enhancement in the ΔH value. The temperature and pressure were kept at 298.15 K and 1.0 atm, respectively, for all the calculations. For the calculation of binding enthalpy (ΔH), the electronic ($E_{elec}$), vibrational ($H_{vib}$), and zero-point vibration energies (ZPE) of the systems were considered i.e.,

$$H = E_{elec} + ZPE + H_{vib} \quad \ldots (3)$$

The energy decomposition analysis (EDA) has been performed using the scheme proposed by Su and Li popularly known as the localized molecular orbital energy decomposition analysis (LMO-EDA) to identify the contribution of different energy terms to the total interaction energy.[45] The geometries obtained at the UB3LYP-D3/cc-pVTZ level were further used to perform the LMO-EDA using the general atomic and molecular electronic structural system "GAMESS" suite code.[46] In the case of LMO-EDA, the total interaction energy ($\Delta E_{int}$ or $\Delta H_{bind}$) of a system contains contribution from electrostatic ($\Delta E_{ele}$), exchange ($\Delta E_{ex}$), repulsion ($\Delta E_{rep}$), polarization ($\Delta E_{pol}$), and dispersion ($\Delta E_{dis}$) energy components. The total interaction energy was calculated using the expression as shown in the equation 4.

$$\Delta H_{bind} = \Delta E_{ele} + \Delta E_{ex} + \Delta E_{rep} + \Delta E_{pol} + \Delta E_{dis} \quad \ldots (4)$$

Also, Frontier Molecular Orbital (FMO) analysis has been performed to find the stability of the complexes showing strong value of ΔH. The geometries obtained at the UB3LYP-D3/cc-pVTZ level of theory was used to calculate the energies of the highest occupied molecular orbital (HOMO) and lowest unoccupied molecular orbital (LUMO). The energy difference or energy gap ($E_g$) between the HOMO and LUMO calculated using equation 5.

$$E_g = E_{LUMO} - E_{HOMO} \quad \ldots (5)$$

Where $E_{HOMO}$ and $E_{LUMO}$ are the energy of the HOMO and LUMO, respectively.

## 3. MATERIALS DESIGN AND ADSORPTION SITES

The main linker used in this study is the basic building unit of the first synthesized MOF material (i.e., 1,4-benzene dicarboxylate in short BDC ligand or linker) with halogen group elements attached to different positions of the BDC linker. The linkers used in this study contains hydrogen (H), carbon (C), oxygen (O), Fluorine (F), and Chlorine (Cl) atoms as their constituent atoms. BDC is basically a benzene ring with a carboxyl group attached at the para positions of the ring, commonly known as 1,4-benzene dicarboxylate. FBDC is 3-fluoro 1,4-benzene dicarboxylate, where one fluorine is attached to the meta position of the BDC linker. DFBDC-1 and DFBDC-2 are difluoro 1,4-benzene dicarboxylate with two fluorine atoms attached to different positions of the BDC linker. In the case of DFBDC-1, one fluorine is attached at the meta and the other is at the ortho position (i.e., 3,6-difluoro 1,4-benzene dicarboxylate) whereas, in the case of DFBDC-2, one fluorine is at the ortho while the other is at the meta position of the BDC ring (i.e., 2,3-difluoro 1,4-benzene dicarboxylate). TFBDC is 2,3,5,6-tetrafluoro 1,4-benzene dicarboxylate ligand which is formed by attaching four fluorine atoms at both the ortho and para positions. DClBDC-1 and DClBDC-2 are dichloro 1,4-benzene dicarboxylate with two chlorine atoms attached to different positions of the BDC linker. In the case of DClBDC-1, one chlorine is attached at the meta position while the other is at the ortho position (i.e., 3,6-chloro 1,4-benzene dicarboxylate) whereas, in the case of DClBDC-2, one chlorine is attached at the ortho while the other is at the meta position (i.e., 2,3-dichloro 1,4-benzene dicarboxylate) of the BDC ring. NDC is 1,4 naphthalene dicarboxylate which is formed by a fused pair of benzene rings with carboxylate groups attached to the para positions. All the links discussed above are shown in Figure 1.

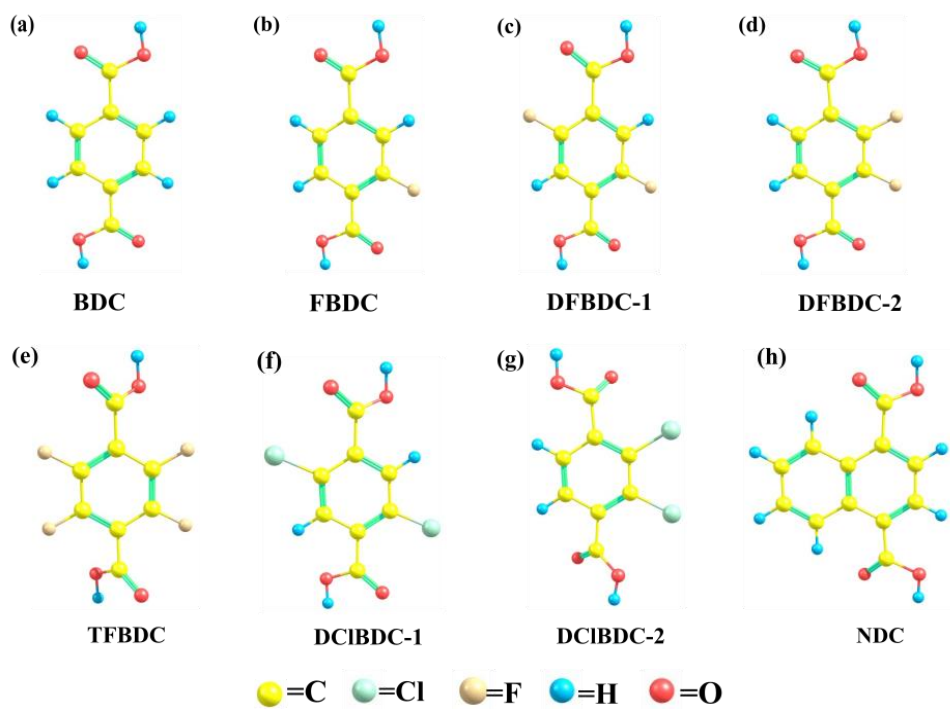

**Figure 1.** The equilibrium structures of the pure MOF's linkers: (a) 1,4-benzene dicarboxylate (BDC), (b) 3-fluoro 1,4-benzene dicarboxylate (FBDC), (c) 3,6-difluoro 1,4-benzene dicarboxylate (DFBDC-1), (d) 2,3-difluoro 1,4-benzene dicarboxylate (DFBDC-2), (e) 2,3,5,6-tetrafluoro 1,4-benzene dicarboxylate (TFBDC), (f) 3,6-dichloro 1,4-benzene dicarboxylate (DClBDC-1), (g) ) 2,3-dichloro 1,4-benzene dicarboxylate (DClBDC-2), and (h) 1,4 naphthalene dicarboxylate (NDC).

Understanding the adsorption mechanism in a porous material requires knowledge of the adsorption sites of the adsorbate molecules, so we chose the center and side positions of the linkers as the adsorption sites. For each system, the $CO_2$ molecules was placed perpendicularly to the center and side position of the pure and Li-decorated linkers. Figure 2 depicts the adsorption of $CO_2$ molecule at the center and side positions of the BDC linker.

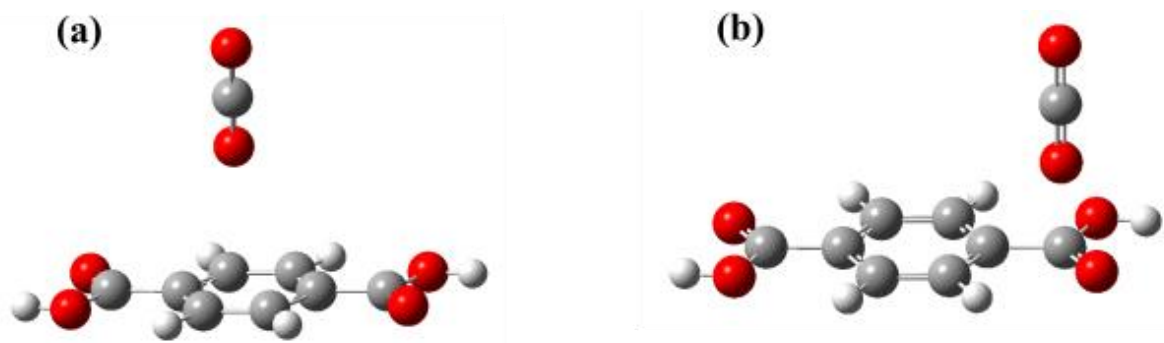

**Figure 2.** The adsorption of CO2 molecules at the (a) Center, and (b) Side position of the BDC linker.

## 4. RESULTS AND DISCUSSIONS

### 4.1 Binding Enthalpy

The change of enthalpy (ΔH) is one of the deciding parameters in determining the stability of any hypothetical compound or complex system. A negative value of ΔH suggests that the structure is thermodynamically stable and can be synthesized experimentally. Here, the adsorption of $CO_2$ molecule at the center and side positions of the pristine and Li-decorated linkers have been investigated. The different adsorption site of $CO_2$ on the linkers can be understood by examining the bonding nature of the linkers with $CO_2$. Firstly, the adsorption of $CO_2$ on all the pristine linkers were studied and the strength of their interactions were investigated using equation 1. It was observed that after optimization, $CO_2$ molecule moved away from its original center and side positions. The equilibrium geometries of the complexes are shown in the Figure 3 and the cartesian coordinates of the optimized structures are provided in the Supporting Information. The computed value of ΔH lies in the range of -8.07 to -14.09 kJ/mol computed by the B3LYP-D3 method, while the value lies between -6.76 to -13.98 kJ/mol for the MP2 method. It is observed that all the calculated values of ΔH were negative, which also indicates the stability and exothermic nature of the adsorption processes. These values of ΔH indicates that $CO_2$ weakly interacts with pristine linkers via van der Waals forces where the π electrons of the benzene rings constituting the linkers attracts the positively charged carbon atom of $CO_2$. Moreover, from these calculations, we found the favourable adsorption sites of $CO_2$. The highest binding configuration is observed for the NDC linkers with one of the O atoms of $CO_2$ pointing towards the center of the six-membered ring, having ΔH value of -14.09 and -13.98 kJ/mol computed by the B3LYP-D3 and MP2 methods respectively. This strong value of ΔH is due to the increased surface area of the NDC linker, as it contains one extra benzene ring as compared to the other linkers. Additionally, it was found that fluorine and chlorine attached to the BDC linker at the ortho, meta, and para positions significantly influence the value of ΔH. $CO_2$ adsorption at the center and side positions of the FBDC, DFBDC-1, DFBDC-2, TFBDC and DClBDC-2 linkers is showing favourable physisorption behaviour.

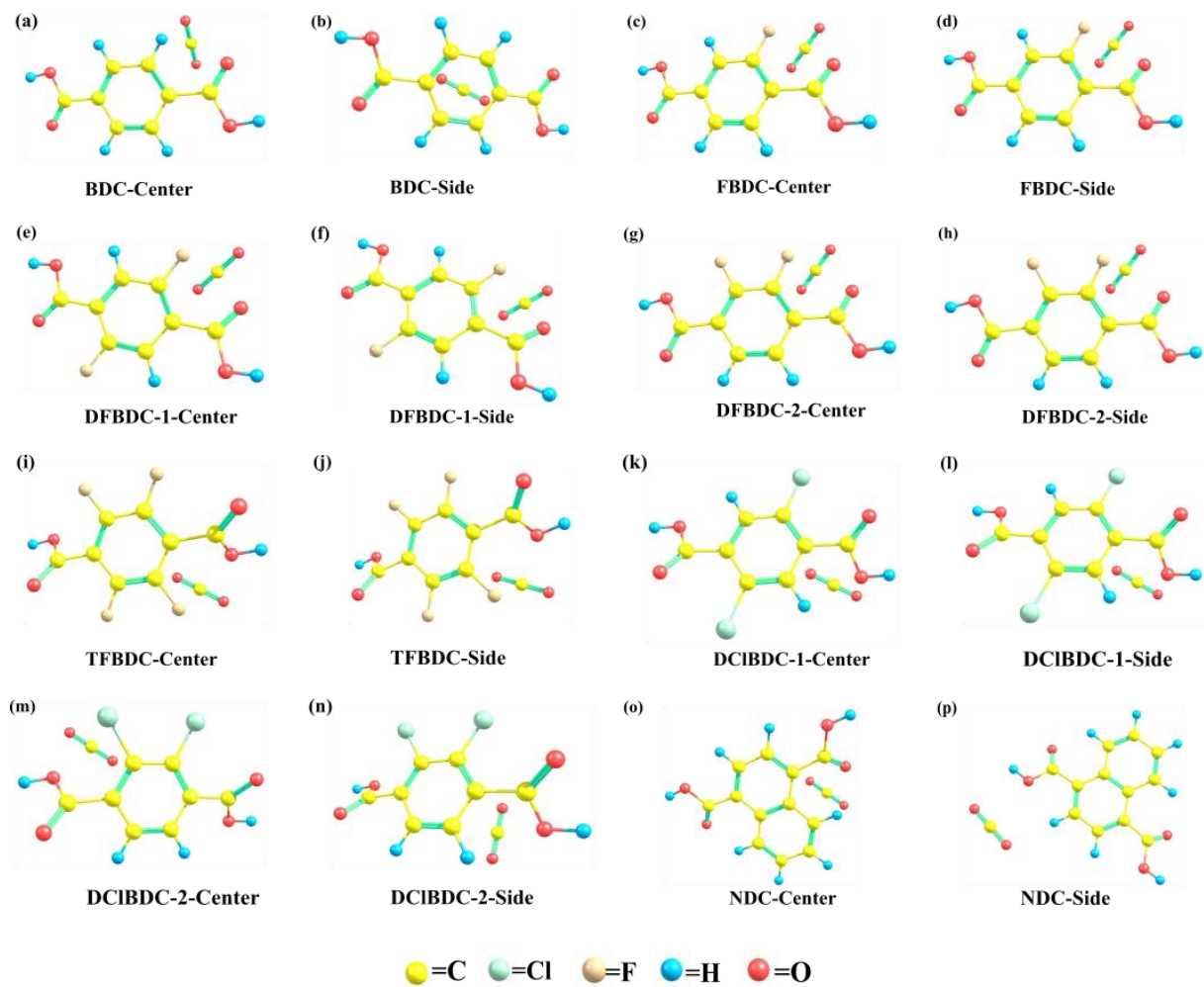

**Figure 3.** Equilibrium structures of both the pure/pristine linkers and the adsorbed $CO_2$ at the center and side positions of the linkers: (a) BDC-Center, (b) BDC-Side, (c) FBDC-Center, (d) FBDC-Side, (e) DFBDC-1-Center, (f) DFBDC-1-Side, (g) DFBDC-2-Center, (h) DFBDC-2-Side, (i) TFBDC-Center, (j) TFBDC-Side, (k) DClBDC-1-Center, (l) DClBDC-2-Side, (m) DClBDC-2-Center, (n) DClBDC-2-Side, (o) NDC-Center, and (p) NDC-Side

Next, we performed Li-decoration to see the effect on the binding strength of $CO_2$ with the Li-decorated complexes. According to Gu et al. a higher negative value of ΔH corresponds to a much stronger binding process.[30] Due to the strong affinity of positively charged metal ions to the adsorbing gases, this strategy has been widely implemented for hydrogen storage.[31,47–50] The adsorption of Li atoms on all the linkers forming Li-decorated complexes was studied by fully optimizing the structures, with Li atoms originally positioned near to the O atoms of the COOH group attached to the six-atom carbon ring ($C_6H_4$ units). After geometry optimization, lithium atom preferred to be in a position between the two neighbouring oxygen atoms. To further investigate $CO_2$ capture, we optimized the structures with $CO_2$ positioned close to the Li-atoms i.e., at the center and side positions as performed for the pristine linkers. The structures were little distorted after geometry optimization, due to charge transfer occurring from the Li atom to the linkers. The equilibrium geometries of the complexes are shown in the Figure 4 and the cartesian coordinates of the optimized structures are provided in the Supporting Information.

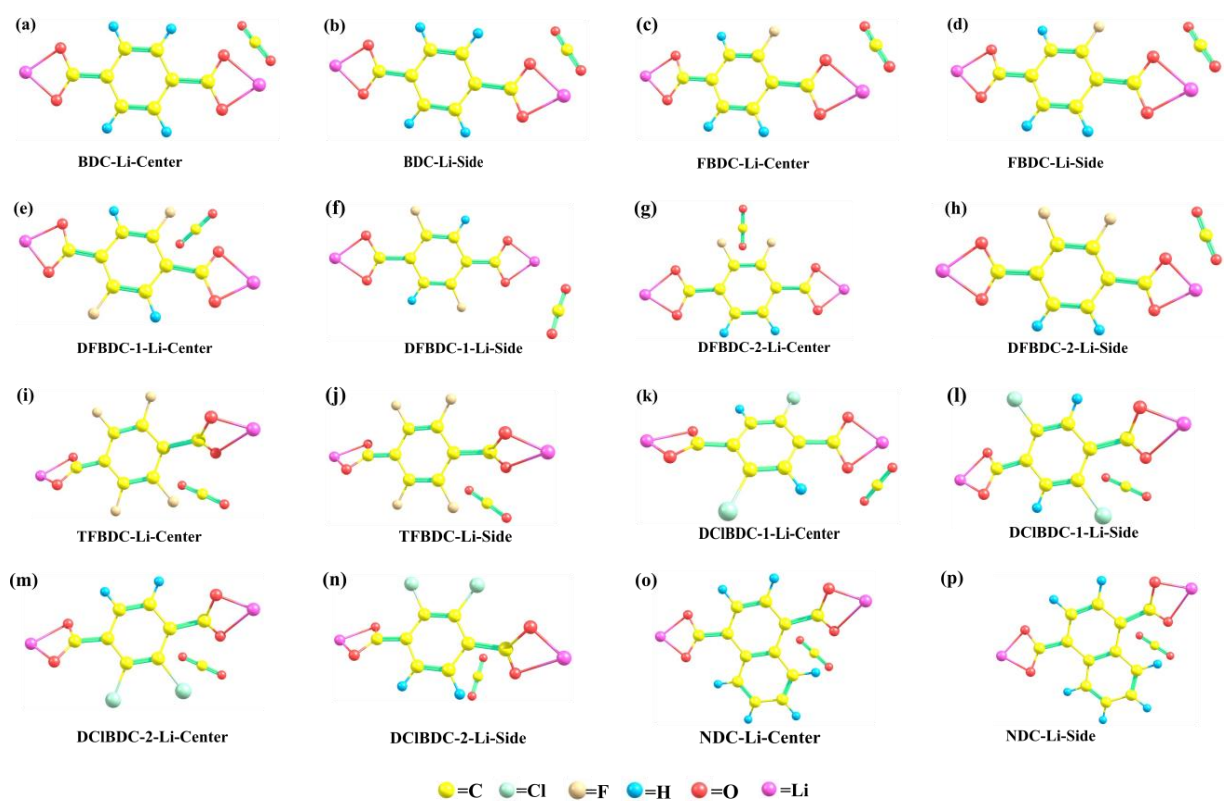

**Figure 4.** Optimized structures of the Li-decorated linkers and the adsorbed $CO_2$ at the center and side positions: (a) BDC-Li-Center, (b) BDC-Li-Side, (c) FBDC-Li-Center, (d) FBDC-Li-Side, (e) DFBDC-1-Li-Center, (f) DFBDC-1-Li-Side, (g) DFBDC-2-Li-Center, (h) DFBDC-2-Li-Side, (i) TFBDC-Li-Center, (j) TFBDC-Li-Side, (k) DClBDC-1-Li-Center, (l) DClBDC-2-Li-Side, (m) DClBDC-2-Li-Center, (n) DClBDC-2-Li-Side, (o) NDC-Li-Center, and (p) NDC-Li-Side

Based on the calculation results, after Li-decoration, $CO_2$ adsorbed on the side position of the DFBDC-2 linker is the most favourable adsorption site having ΔH value of -35.32 kJ/mol computed by the B3LYP method which is almost three times the value of pristine DFBDC-2. This value of Li-decorated DFBDC-2 linker shows that Li contributes an enthalpy of about -22.03 kJ/mol to the total value of ΔH. For the MP2 method, the ΔH value for DFBDC-2 was found to be -34.13 kJ/mol in Li-decorated case, while -11.96 kJ/mol for pure one, which is more than three times the value observed in Li-decorated DFBDC-2. It was observed that the ΔH values computed by the MP2 method are comparable with the values achieved by the B3LYP-D3 method. This high affinity of $CO_2$ with the linkers is due to the high polarizability and quadrupole moment of $CO_2$. It is found from previous studies that electrostatic interactions between the quadrupolar $CO_2$ and the lone pairs of the carbonyl oxygen of the carboxyl group contribute to this kind of increment in binding enthalpy.[22] The values of ΔH for $CO_2$ adsorbing at the center and side positions of the linkers computed by the DFT B3LYP-D3 and MP2 method are shown in Figure 5 and 6 respectively, and the calculated values of ΔH for the pristine and Li-decorated linkers are provided in Table S1 and S2 of the Supporting Information. It was observed from Figure 5 and 6 that both the positions of the BDC and FBDC linkers have large ΔH value, and the same trend can be seen for the side position of the DFBDC-1, DFBDC-2 and DClBDC-1 computed by both the methods. This increase in the value of ΔH for FBDC, DFBDC-1, DFBCD-2, and DClBDC-1 linkers is due to the nature of the electron-withdrawing group of fluorine and chlorine atoms. These values are comparable with the results reported for the effect of $SO_3H$ functional group for $CO_2$ capture on MOF linkers.[30] The adsorption of $CO_2$ at the center and side positions of the TFBDC, DClBDC-2, and NDC lies in the range of -15.02 kJ/mol to -17.45 kJ/mol computed by the B3LYP-D3 method and -14.86 kJ/mol to -16.45 kJ/mol for the MP2 method. These values are comparable with the previous reported results for the functionalized linkers studied for $CO_2$ capture on MOF and ZIF linkers.[30,51] It is interesting to see that $CO_2$ adsorption on the NDC linker does not show a large increment in ΔH value as was observed for the other linkers after Li decoration. The results from the harmonic vibrational analysis of $CO_2$ adsorbing on the pristine and Li-decorated linkers shows no imaginary frequency, indicating that all these structures correspond to the local minima on the potential energy surface. In other words, the absence of imaginary frequencies assures the minima on the potential energy surface. Also, the negative value of ΔH indicates that the practical formation of porous materials synthesized using these linkers can be possible. These linkers show strong affinity with the $CO_2$ molecule, and hence can be used as an efficient adsorbent for the future designing of an ideal porous material.

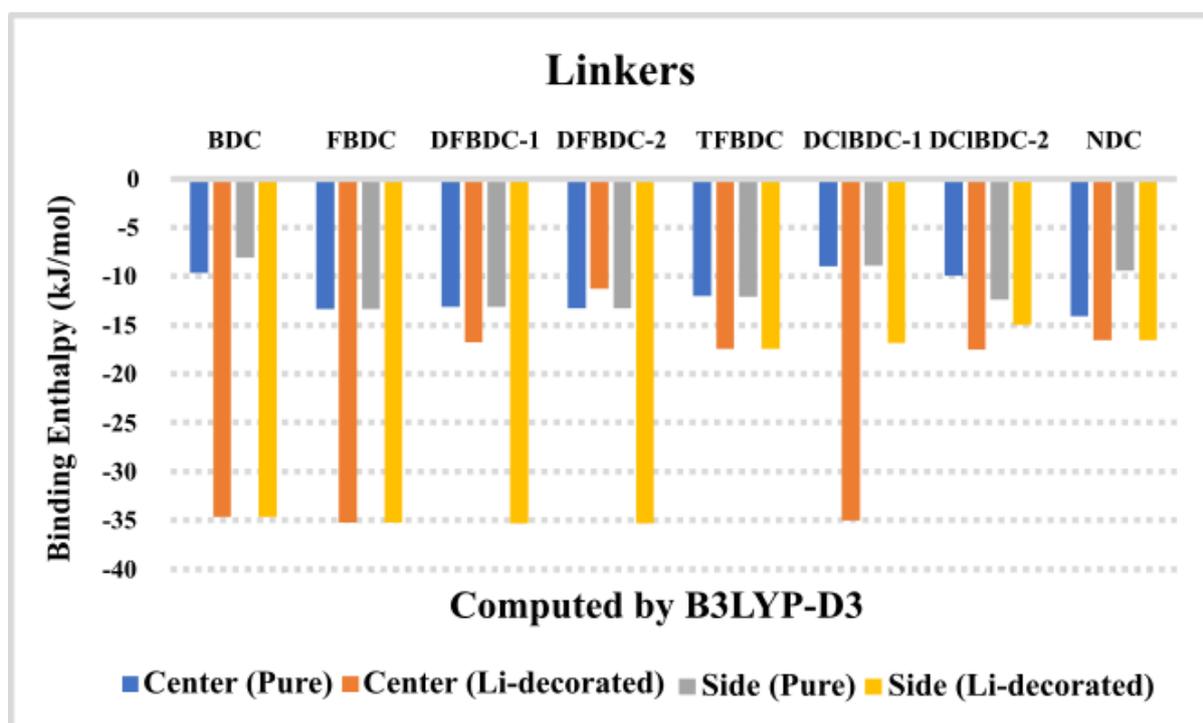

**Figure 5.** The binding enthalpies (ΔH) of the adsorbed CO$_2$ molecule at the center and side positions of the pure and Li-decorated linkers computed by the DFT B3LYP-D3 method.

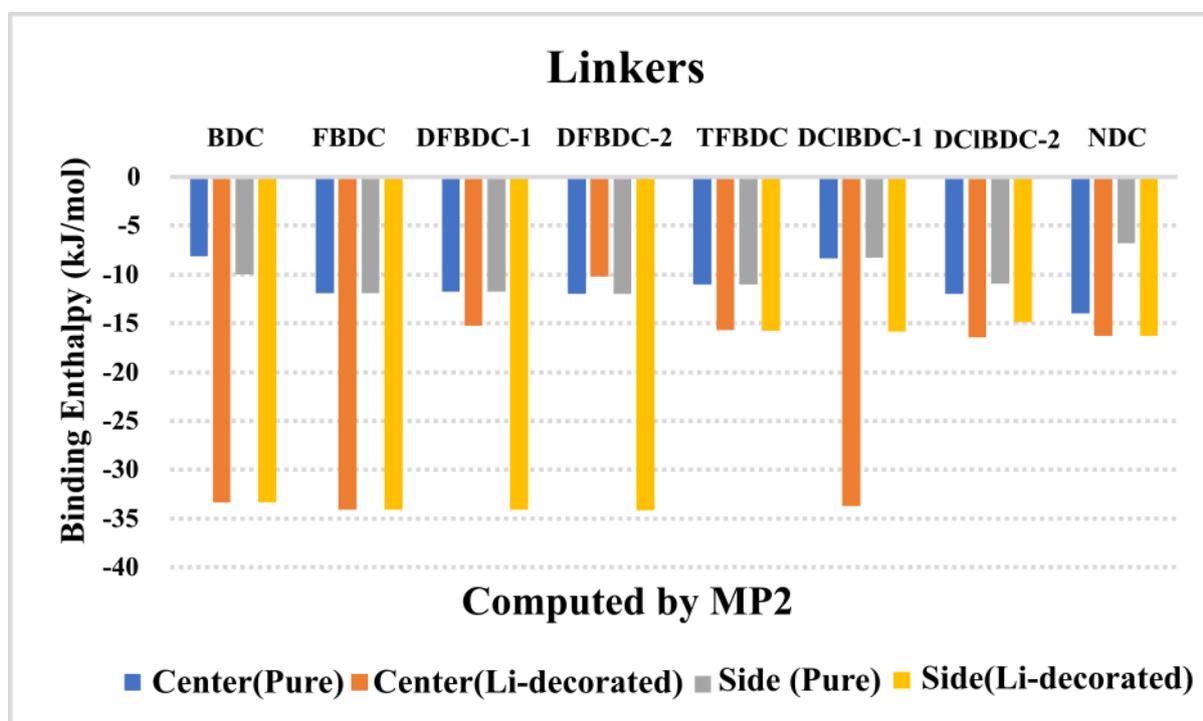

**Figure 6.** The binding enthalpies (ΔH) of the adsorbed CO$_2$ molecule at the center and side positions of the pure and Li-decorated linkers computed by the MP2 method.

Based on the above discussion of ΔH, it was found that the side positions of the BDC, FBDC, DFBDC-1, and DFBDC-2 shows strong affinity with $CO_2$, while the other linkers show the physisorption behaviour of $CO_2$ adsorption on the linkers. For reversible adsorption, the linkers showing physisorption behaviour can be used as an efficient adsorbent in designing MOF materials for the adsorption of gases.

## 4.2 Energy Decomposition Analysis (EDA)

EDA is one of the quantitative methods based on electronic structure calculations that tells about the contribution of various energy terms to the total interaction energy. It provides understanding of the factors involved in molecular interactions. The result of this analysis helps us to predict the properties of an interacting molecular system which are difficult to analyze experimentally. Here, EDA has been performed for the $CO_2$ molecule getting adsorbed to the side position of the BDC, FBDC, DFBDC-1 and DFBDC-2 linker. The contribution from different energy terms to the total ΔH value of most stable linkers are shown in the Figure 7. Our study shows that electrostatic, exchange, repulsion, polarization, and dispersion contribute energy to the total binding enthalpy. Through this EDA, it is found that these complexes are effectively stabilized by electrostatic as well as polarization components or in order words, the study reveals that the interaction between the Li-decorated linkers and $CO_2$ molecule arises mainly from the electrostatic and polarization energy terms. The contribution from exclusion and dispersion energy components to ΔH value was also found to be considerable. The repulsion energy term shows a positive contribution to the energy which is obvious due to Pauli principle. Thus, it is observed that among all the energy terms, the electrostatic energy term is the most dominant one.

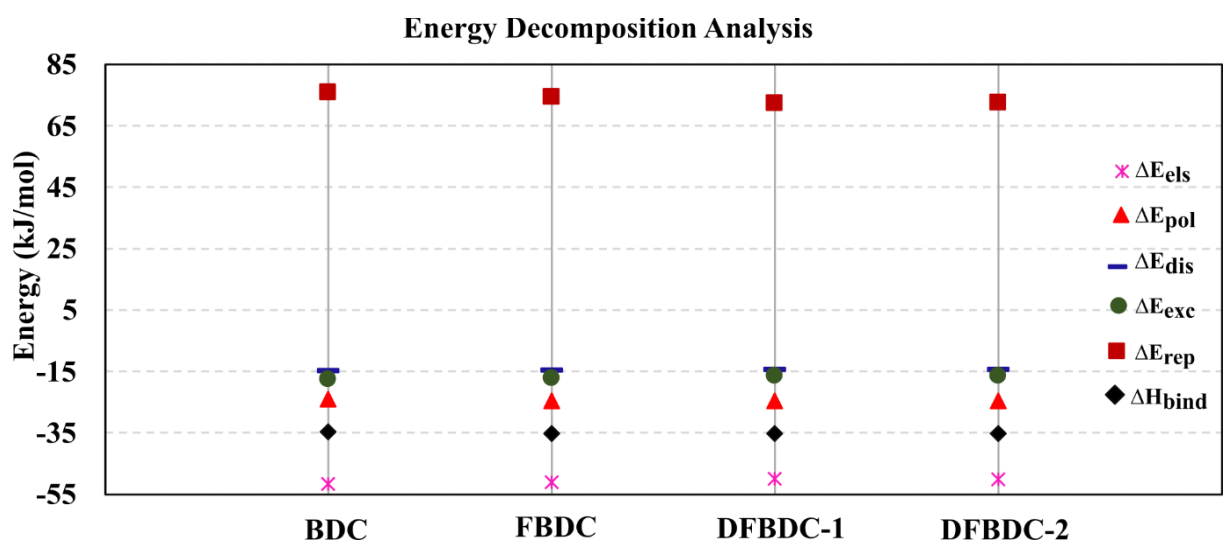

**Figure 7.** The Energy Decomposition Analysis (EDA) of $CO_2$ interacting at the side position of the linkers. The electrostatic energy, polarization energy, dispersion energy, exchange energy, repulsion energy, and binding enthalpy are represented by $\Delta E_{els}$, $\Delta E_{pol}$, $\Delta E_{dis}$, $\Delta E_{exc}$, $\Delta E_{rep}$, and $\Delta H_{bind}$ respectively. The units are expressed in kJ/mol.

### 4.3 Frontier Molecular Orbital (FMO)

To provide additional support to our results and understand the interaction between the orbitals of adsorbed $CO_2$ and the organic linkers, we extend our study to the overlapping of molecular orbitals. The frontier molecular orbital analysis (FMO) was used for determining the overlapping as well as the energies of the highest occupied molecular orbital (HOMO) and lowest unoccupied molecular orbital (LUMO) levels. The HOMO and LUMO are collectively known as Frontier Molecular Orbital (FMO).[52] The difference in energies of HOMO-LUMO gives an energy gap ($E_g$) which explains the stability as well as charge transfer taking place within the complex.[53] The high value of $E_g$ indicates the high kinetic stability and low chemical reactivity of the complex and hence are chemical harder.[54,55]

This analysis was applied to all the complexes for $CO_2$ molecule getting adsorbed at the side positions of the linkers. For both pure and Li-decorated linkers, we investigated the $E_g$ value by calculating the energy of the HOMO and LUMO for all the linkers with $CO_2$ adsorbing at the side position of the linkers. The computed energies of HOMO and LUMO of all the linkers and the energy gap between them are provided in the Supporting Information. It is observed from the Figure 8 that there is a very small difference between the energy gap of FMO in the pure and Li-decorated linkers and the energy gaps of the Li-decorated linkers are higher than the pure linkers, proving that the complexes remained stable even after Li-decoration. The contribution from the $CO_2$ molecules in the HOMO and LUMO calculation of the complexes alone is -10.34 and 0.736 eV respectively. The $E_g$ value of the most stable linkers i.e., BDC, FBDC, DFBDC-1 and DFBDC-2 are 5.33eV, 5.09eV, 4.83 eV and 4.97 eV respectively. Among all the complexes showing strong binding affinity or $\Delta H$, the highest HOMO-LUMO gap (5.33 eV) is observed for the BDC linker where the HOMO and LUMO are -6.69 and -1.37 eV respectively. The lowest value of HOMO-LUMO is observed for DFBDC-1 linker having a gap of 4.83 eV. The chances of electrons going from HOMO to LUMO decrease as the difference between the energies of FMO increases. These high values of the HOMO-LUMO gap for BDC, FBDC, DFBDC-1, and DFBDC-2 demonstrate higher kinetic stability of the complexes.

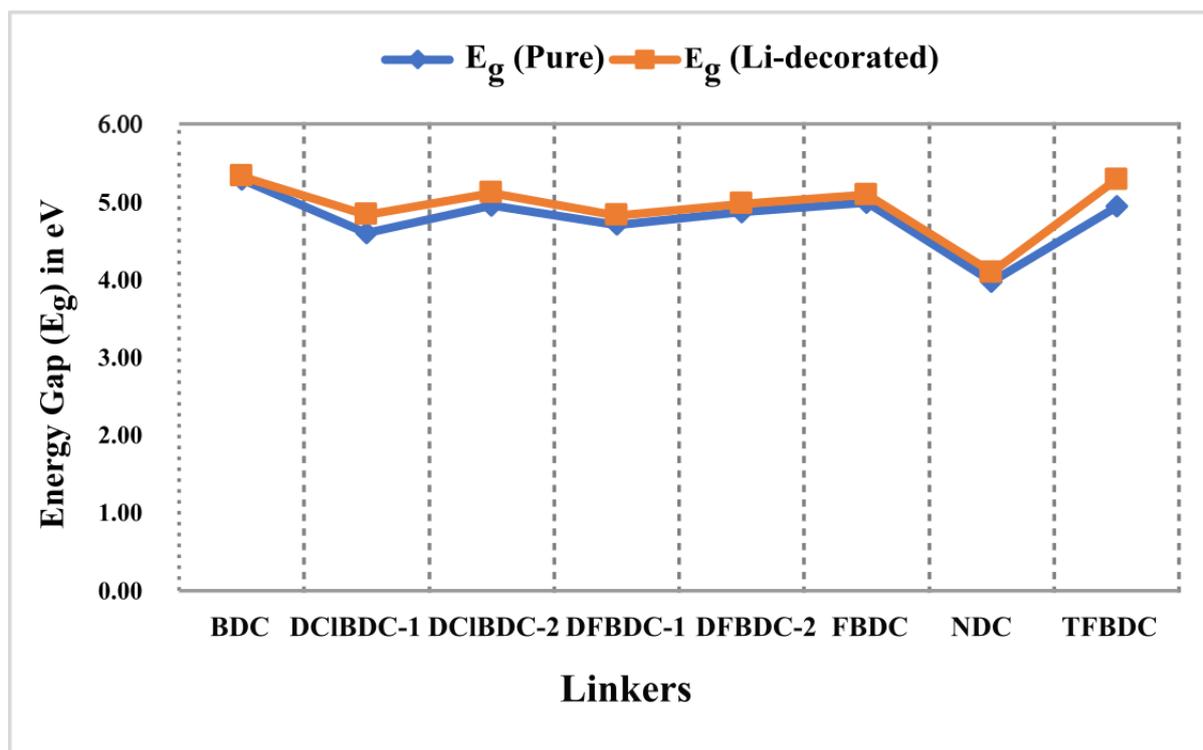

**Figure 8.** The energy gap ($E_g$) between the HOMO and LUMO of the pure and Li-decorated linkers. The energies are expressed in eV.

The Figure 9 shows the overlapping of the orbitals of the pure and Li-decorated linkers with the $CO_2$ molecule. Figure 9 (a)-(d) shows the orbital overlappings of pure linkers with the adsorbed $CO_2$ molecule, where interaction is taking place between the p orbitals of C, O and F atoms and $CO_2$ molecule. Similarly, Figure 9 (e)-(h) shows the overlapping of Li-decorated linkers, where the p orbitals of Li, C, O and F atoms are interacting with the p orbital of $CO_2$ molecule. The positive wave function i.e. up spin of the electrons or alpha electron is indicated by the red colour orbital, while the negative wave function i.e. down spin of the electrons or beta electron is indicated by the blue colour orbital. It is observed from the Figure 9 that the p orbitals of the Li, C, O and F atoms in the pure and Li-decorated linkers are interacting with the p orbital of the $CO_2$ molecule. These molecular interaction shows that $CO_2$ molecule is strongly adsorbed i.e., physisorbed on the side position of the BDC, FBDC, DFBDC-1 and DFBDC-2 linkers.

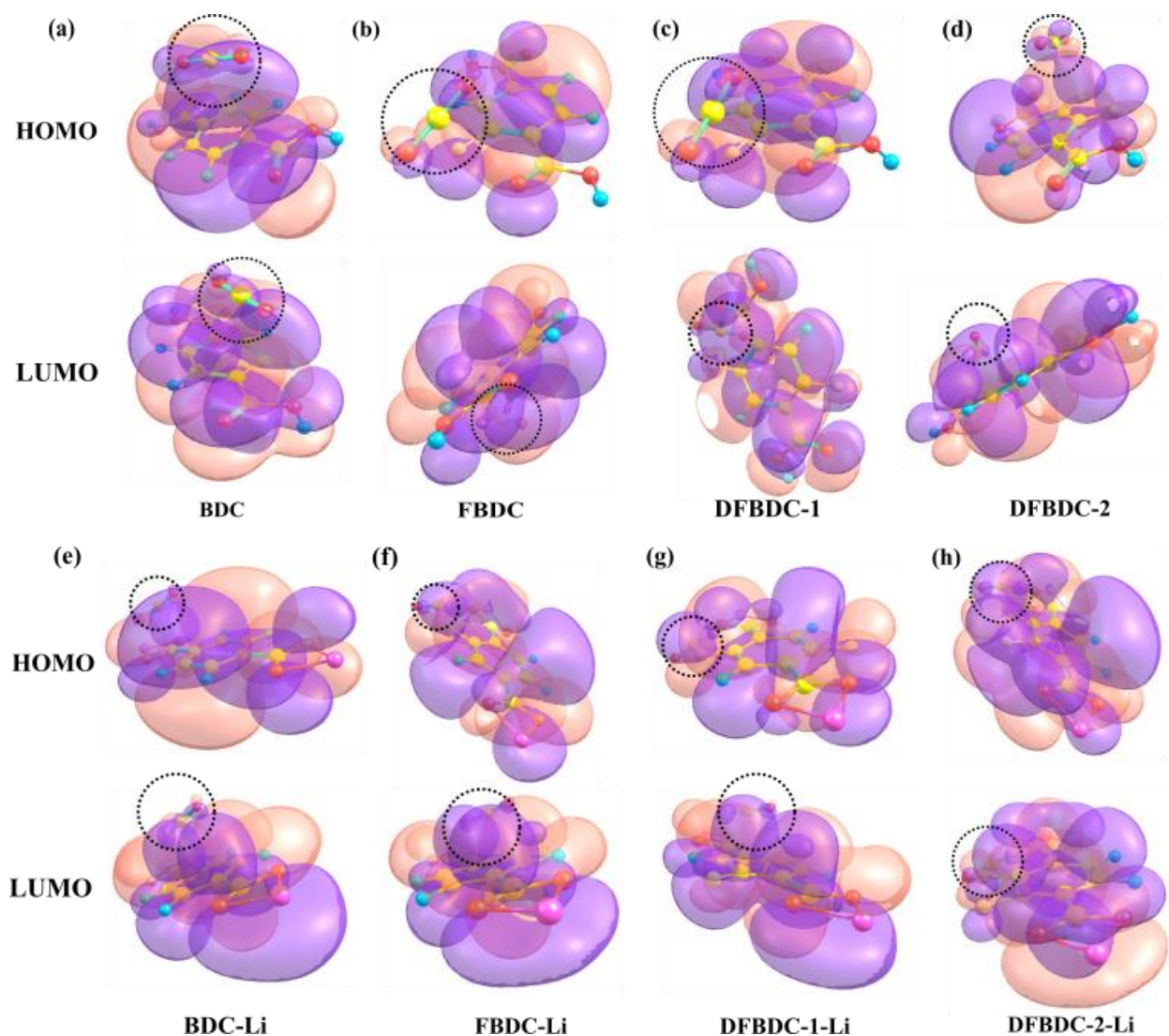

**Figure 9.** The HOMO-LUMO projections of $CO_2$ at the side position of pure and Li-decorated linkers (a) BDC linker, (b) FBDC linker, (c) DFBDC-1 linker, (d) DFBFC-2 linker, (e) BDC-Li-decorated linker, (f) FBDC-Li-decorated linker, (g) DFBDC-1-Li-decorated linker, and (h) DFBDC-2-Li-decorated linker.

It is clear from the above discussion that among all the stable complexes, BDC has the highest $E_g$ value of 5.29 and 5.33 eV for the pure and Li-decorated case respectively. The $E_g$ value for the FBDC, DFBDC-1 and DFBDC-2 are 4.99 eV, 4.70 eV, 4.87 eV for the pure case, while for Li-decorated case the values are 5.09 eV, 4.83 eV and 4.97 eV respectively. These higher value of $E_g$ indicate the kinetic stability of the complexes.

## 5. CONCLUSIONS

In this study, we have computationally explored the impact of decorating light electropositive element i.e., Li on the linkers of the MOF materials and their interaction with the $CO_2$ molecule using molecular model systems. The UDFT method with the cc-pVTZ basis set has been used to provide a quantitative insight into the adsorption of $CO_2$ gas molecule on the linkers of the MOF materials at the atomistic level. The strength of interaction between the complex has been calculated using the DFT-B3LYP and MP2 methods, and the interaction is found stronger for the B3LYP-D3 method. MP2 was found giving a very closer result to B3LYP-D3. Among all the linkers, the most stable structure is DFBDC-2 with $CO_2$ getting adsorbed at the side position of the linkers having an enthalpy of about -35.32 kJ/mol after Li-decoration which is three times larger than the ΔH value of the pristine linker (i.e., 13.29 kJ/mol) computed by the B3LYP-D3 method. An approximate value and similar trend are also observed for the BDC, FBDC, and DFBDC-1 linker with high increase in the value of ΔH after Li-decoration. These results indicate that metal decoration is an efficient approach in designing linkers with highly improved $CO_2$ affinity via physisorption. Furthermore, the results from the EDA reveals that the contribution from electrostatic and polarization energy terms to the ΔH value are the most dominant one. The energy difference between the HOMO and LUMO reveals that the linkers BDC, FBDC, DFBDC-1 and DFBDC-2 have $E_g$ values 5.33eV, 5.09eV, 4.83 eV and 4.97 eV respectively, thereby indicating the stability of the complexes even after Li-decoration. To get a deep understanding of the interaction mechanism, the value of ΔH plays a crucial role in the designing of an ideal adsorbent material for the adsorption of gases. We expect that the results of our study represent significant and encouraging steps towards the development novel $CO_2$ capture adsorbent materials in the near future.

**SUPPORTING INFORMATION**

The supporting information contains binding enthalpy of the pristine and Li-decorated linkers at the center and side positions of the linkers, energies of the HOMO and LUMO for CO2 adsorption at the side positions of all the linkers, all the coordinates of optimized structure including linkers, pure and Li-decorated linkers after $CO_2$ adsorption at the center and side positions.

**AUTHOR INFORMATION**

**Corresponding Author**

Dr. Srimanta Pakhira – *Theoretical Condensed Matter Physics & Advanced Computational Materials Science Laboratory, Department of Physics, Indian Institute of Technology Indore (IIT Indore), Simrol, Khandwa Road, Indore-453552, Madhya Pradesh, India.*

*Department of Metallurgy Engineering and Materials Science (MEMS), Indian Institute of Technology Indore (IIT Indore), Simrol, Khandwa Road, Indore, Madhya Pradesh 453552, India.*

*Centre for Advanced Electronics (CAE), Indian Institute of Technology Indore (IIT Indore), Simrol, Khandwa Road, Indore, Madhya Pradesh 453552, India.*

ORCID: orcid.org/0000-0002-2488-300X

E-mail: spakhira@iiti.ac.in or spakhirafsu@gmail.com

**Author**

Ms. Kahkasha Parveen − *Theoretical Condensed Matter Physics & Advanced Computational Materials Science Laboratory, Department of Physics, IIT Indore, Simrol, Khandwa Road, Indore, Madhya Pradesh 453552, India.*

ORCID: orcid.org/0000-0003-2513-9634

**ACKNOWLEDGEMENTS:**

We thank the Science and Engineering Research Board-Department of Science and Technology (SERB-DST), Government of India, for their financial and technical assistance under Grant Nos. CRG/2021/000572 and ECR/2018/000255. This research project has received funding from the SERB-DST, Government of India, under Grant Nos. SB/S2/RJN-067/2017 ECR/2018/000255 and CRG/2021/000572. We appreciate the Core Research Grants (CRG) provided by the SERB-DST under the CRG/2021/000572 scheme. Dr. Srimanta Pakhira expresses gratitude to his Early Career Research Award (ECRA) under grant number ECR/2018/000255 and the Science and Engineering Research Board (SERB-DST), Government of India, for granting his highly esteemed Ramanujan Faculty Fellowship under scheme number SB/S2/RJN-067/2017. Ms. Kahkasha Parveen thanks INSPIRE, SERB-DST, Govt. of India, for providing her doctoral Inspire fellowship under scheme no. IF200014.

**Conflicts of Interest:**

The authors have no conflicts of interest.

**GRAPHICAL ABSTRACT**

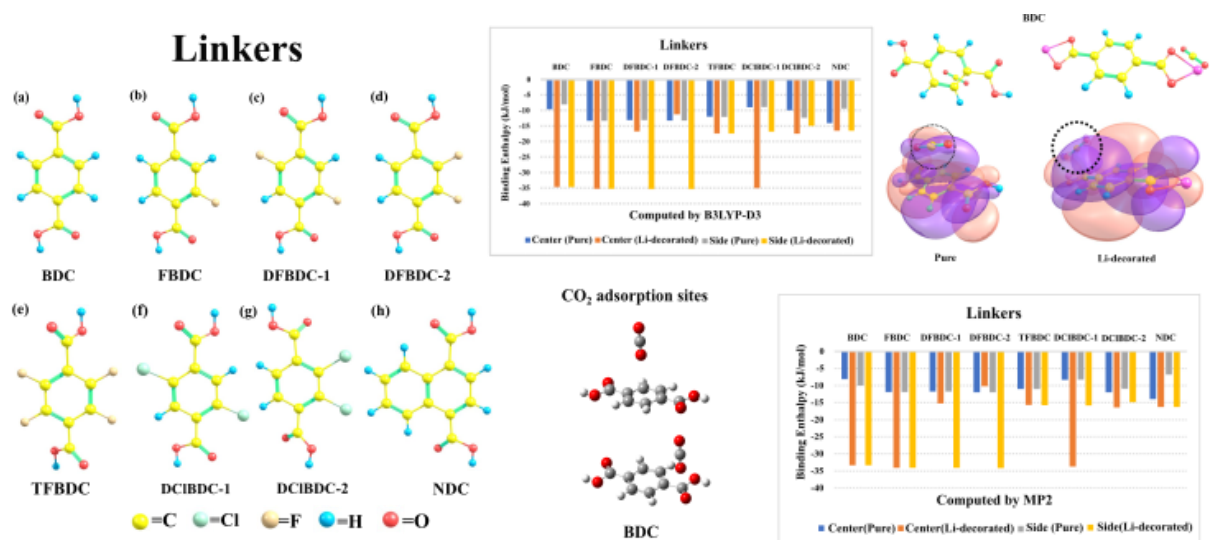